\documentclass[twocolumn,showpacs,preprintnumbers,prl]{revtex4}
\usepackage{amssymb}

\usepackage{graphicx}
\usepackage{dcolumn}
\usepackage{bm}
\usepackage[usenames]{color}

\begin{document}


\title{Spontaneous Four-Wave Mixing of de Broglie Waves: Beyond Optics}
\author{V.~Krachmalnicoff$^{1}$, J.-C.~Jaskula$^{1}$, M. Bonneau$^{1}$,
V. Leung$^1$ G.~B.~Partridge$^{1}$,
D.~Boiron$^{1}$, C.~I.~Westbrook$^{1}$, P.~Deuar$^2$, P.~Zi\'n$^3$,
M.~Trippenbach$^{4}$, and K.V.~Kheruntsyan$^5$}
\affiliation{
$^1$Laboratoire Charles Fabry de l'Institut d'Optique, Univ Paris
Sud, CNRS,
Campus Polytechnique RD128 91127 Palaiseau France\\
$^2$Institute of Physics, Polish Academy of Sciences, Al. Lotnik\'{o}w
32/46, 02-668 Warsaw, Poland\\
$^3$The Andrzej So{\l }tan Institute for Nuclear Studies,
Ho\.{z}a 69, PL-00-681 Warsaw, Poland\\
$^4$Institute of Theoretical Physics, Physics Department, University
of Warsaw,
Ho\.{z}a 69, PL-00-681 Warsaw, Poland\\
$^5$ARC Centre of Excellence for Quantum-Atom Optics, School of Mathematics
and Physics, University of Queensland, Brisbane, Queensland 4072, Australia}
\date{\today}

\begin{abstract}
We investigate the atom-optical analog of degenerate four-wave
mixing of photons by colliding two Bose-Einstein condensates (BECs) of
metastable helium and measuring the resulting momentum distribution
of the scattered atoms with a time and space resolved detector.  For
the case of photons, phase matching conditions completely define the
final state of the system, and in the case of two colliding
BECs, simple analogy implies a spherical momentum
distribution of scattered atoms.  We find however, that the final
momenta of the scattered atoms instead lie on an ellipsoid whose radii
are smaller than the initial collision momentum.
Numerical and analytical calculations agree with the measurements,
and reveal the interplay between many-body
effects, mean-field interaction, and the anisotropy of the source
condensate.
\end{abstract}

\maketitle



The field of atom optics has developed to the point that one can now
speak of the beginning of ``quantum atom
optics''~\cite{Meystre-book} in which atoms are manipulated in ways
similar to photons and in which quantum fluctuations and
entanglement play an important role.
The demonstration of atom pair production \cite{greiner:05,perrin:07},
either from the dissociation of ultra-cold molecules, a process
analogous to parametric down-conversion
\cite{opatrny:01,kheruntsyan:02,savage:06}, or from collisions of BECs
~\cite{deng:99,vogels:02,vogels:03,davidson}, analogous to
four-wave mixing
(FWM)~\cite{duan:00,pu:00,band:00,bach:02,zin:05,norrie:05,deuar:07,molmer:08,chwedenchuk:08,perrin:08,
ogren:09}, holds considerable promise for generating atomic squeezed
states and demonstrating nonlocal Einstein-Podolsky-Rosen (EPR)
correlations
\cite{opatrny:01,kheruntsyan:02,kheruntsyan:05,ferris:09}. In both
these systems, atom-atom interactions play the role of the nonlinear
medium that allows conversion processes. Atoms are not,
however, exactly like photons, and in spite of their formal
similarity, the processes of pair production of photons and of atoms
exhibit some interesting and even surprising differences that must
be understood in order for the quantum atom optics field to advance.
In this work, we discuss one such effect.

In optical FWM or parametric down conversion \cite{scully:97},
energy conservation requires that the sum of the energies of the outgoing
photons be fixed by the energy of the input photon(s). Phase
matching requirements impose constraints on the directions and
values of the individual photon momenta. A simple case is
degenerate, spontaneous FWM (i.e. two input photons of equal energy)
in an isotropic medium, for which energy conservation and phase
matching require that the momenta of the output photons lie on a
spherical shell whose radius is that of the momenta of the input
photons.


We have performed the atom optical analog of degenerate FWM in
colliding BECs while paying careful attention to the momenta of the
outgoing atoms. We find that unlike the optical case, the output
momenta do \textit{not} lie on a sphere, but rather on an ellipsoid
with short radius \textit{smaller} than the input momentum.
This behavior is due to a subtle combination of
atom-atom interactions, which impose an energy cost for pair
production, and the anisotropy of the condensates, which affects
the scattered atoms as they leave the interaction
region.

Although an analogous effect could exist in optics, optical
nonlinearities are typically so small that the effect is negligible.
However in the process of high-harmonic generation in intense laser fields,
a similar effect has been discussed \cite{balcou:97}. There,
phase matching conditions can become significantly intensity dependent,
and the ponderomotive acceleration of electrons alters
the phase and energy balance of the harmonic generation process.
Thus the ponderomotive force plays a role loosely analogous to that of the
mean-field repulsion in our problem.

To fully understand the results, we have simulated the BEC
collision using a fully quantum, first-principles numerical
calculation based on the positive-$P$ representation
method~\cite{deuar:07,perrin:08}, and find quantitative agreement
with the experiment. We have also analyzed the problem using a
stochastic implementation of the Bogoliubov approach, which allows
us to identify and illustrate the contributions of various
interaction effects in the process.

The experimental setup is similar to that described
in~\cite{perrin:07}. We start from a BEC of $\sim10^5$ atoms
magnetically trapped in the $m_x=1$ sublevel of the $2^{3}S_{1}$
metastable state of helium-4. The trap is cylindrically symmetric
with axial and radial frequencies of $47$~Hz and $1150$~Hz,
respectively. The bias field of $\sim0.25$~G along the $x$-axis
defines the quantization axis.

\begin{figure}[tbp]
\begin{center}
\includegraphics[width=3.95cm]{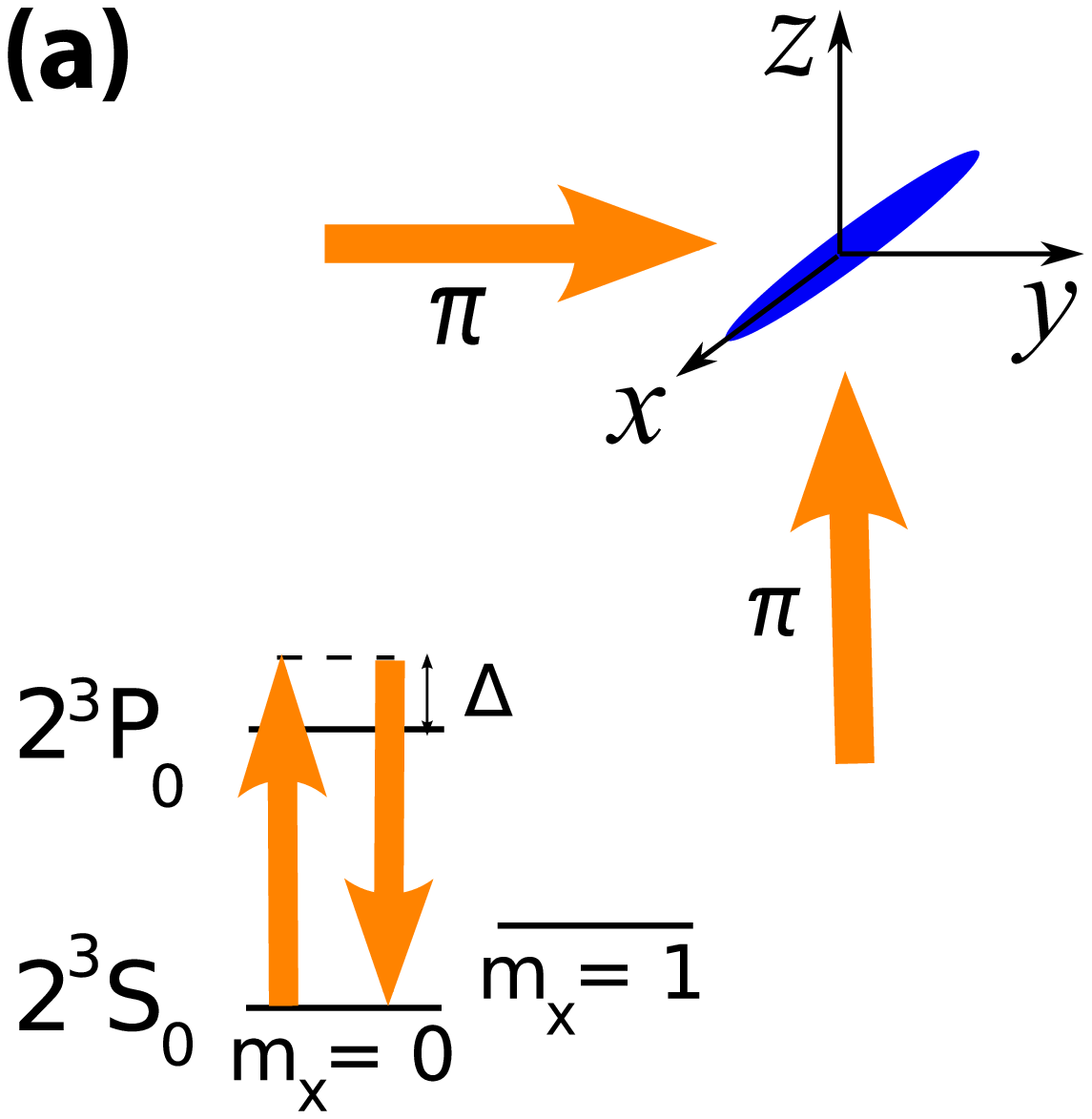}~~
\includegraphics[width=4.45cm]{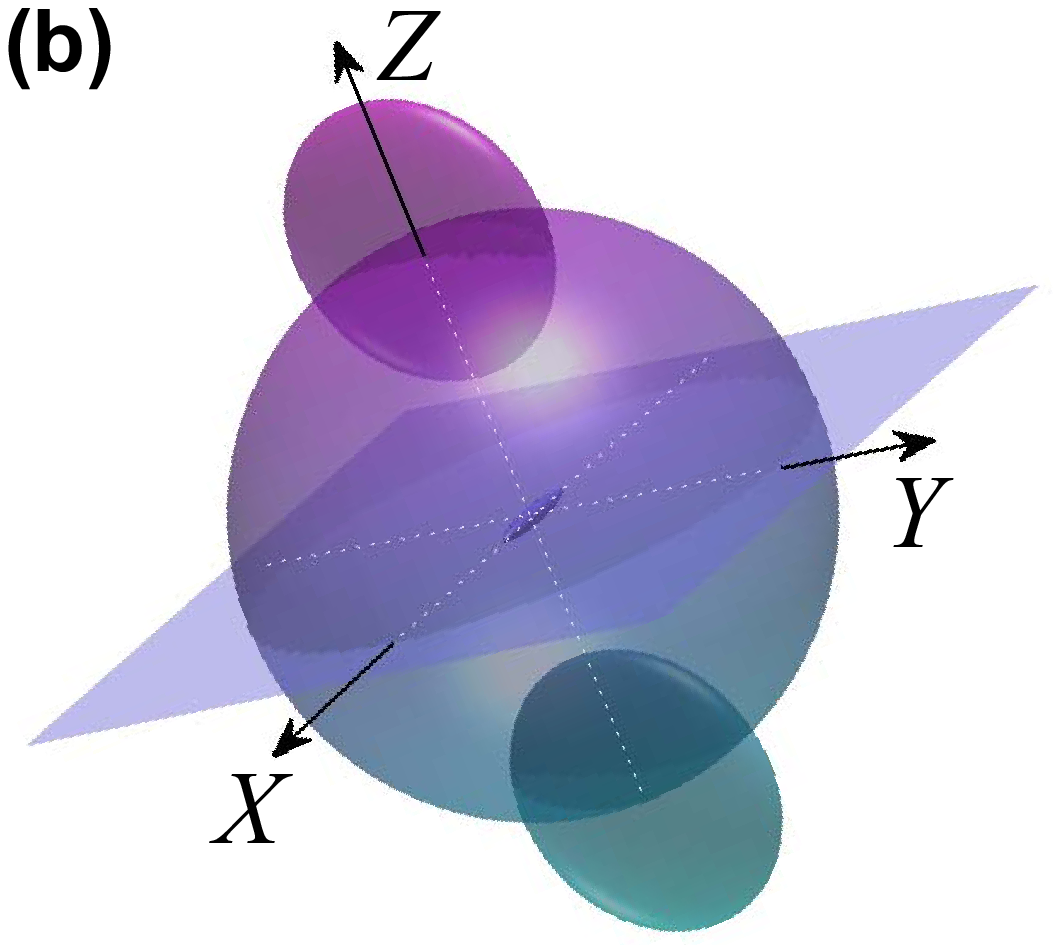}
\caption{\label{fig:beamGeometry} (Color online) (a) Geometry of the
Bragg beams and level scheme of the $2^3 S_1 - 2^3 P _0$ transition
of $^{4}$He (at 1083 nm). A Bragg pulse of two $\pi$-polarized laser beams (shown
by the two arrows) detuned by $\Delta/2\pi=600$~MHz produces two
counterpropagating BECs that separate along their radial
dimension  at approximately $45^\circ$
 to the vertical ($z$) axis at relative velocity $2v_0$.
(b) Schematic diagram of the collision geometry in
the center-of-mass frame in which we denote the collision axis as $Z$.
The two disks represent the colliding condensates in momentum space. The
sphere represents the halo of scattered atoms. The cigar shaped
initial condensate with axial direction $X=x$ is shown in the
center. We analyze the experimental data in the $XY$-plane.
}
\end{center}
\end{figure}

To generate the two colliding BECs, we use a two-step
process. First, the atoms are transferred to the $m_x=0$ state by a
stimulated Raman transition. Using a $4$~$\mu$s long pulse, we transfer
$90$\%~of the atoms to this magnetically untrapped state.
$1$~$\mu$s after the end of the Raman pulse, the BEC is split
into two counterpropagating condensates with a Bragg pulse driven by
two laser beams propagating at approximately $90^\circ$, as shown in
Fig.~\ref{fig:beamGeometry}~(a). The parameters of the Bragg pulse
are adjusted to transfer half of the atoms to a state moving at
relative velocity $2v_0$ in the $yz$-plane, with $v_0=7.31$~cm/s,
which is $\sim4$ times the speed of sound in the center of the
BEC. The condensates thus separate along the \emph{radial}
axis, unlike in the experiment of Ref.~\cite{perrin:07}. To analyze
the data we will use a center-of-mass reference frame, in which the
collision axis is defined as $Z$ (tilted by about $45^\circ$ from
$z$), $X\equiv x$, and $Y$ is orthogonal to $Z$ and $X$
(see Fig.~\ref{fig:beamGeometry}).

After the collision, the atoms fall onto a microchannel plate
detector placed $46.5$ cm below the trap center.
A delay line anode permits reconstruction of a 3D image of the cloud
of atoms. The flight time to the detector ($300$ ms), is long enough
that the 3D reconstruction gives a 3D image of the velocity
distribution after the collision. Binary, $s$-wave collisions
between atoms in the BECs should (\emph{naively}) result in
the scattered particles being uniformly distributed on a sphere in
velocity space with radius equal to the collision velocity $v_0$.
The collision along the radial axis allows access to the entire
collision halo in a plane containing the anisotropy of the
BEC (the $XY$-plane) without distortion from the condensates.
As in Ref.~\cite{perrin:07}, we observe a strong correlation between
atoms with opposite velocities confirming that the
observed halo is indeed the result of binary collisions.

\begin{figure}[tbp]
\begin{center}
\includegraphics[height=4.2cm]{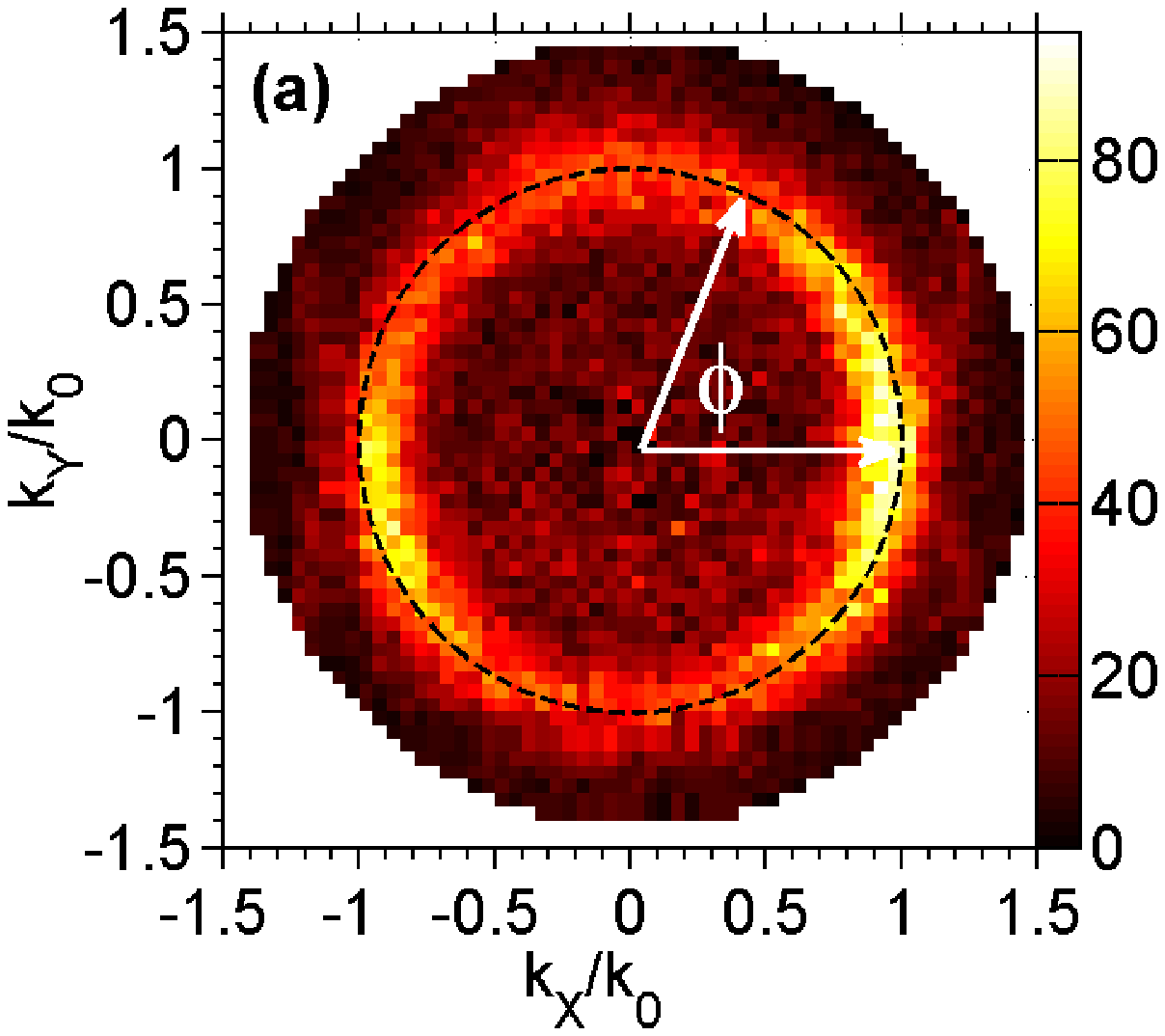}\includegraphics[height=4.2cm]{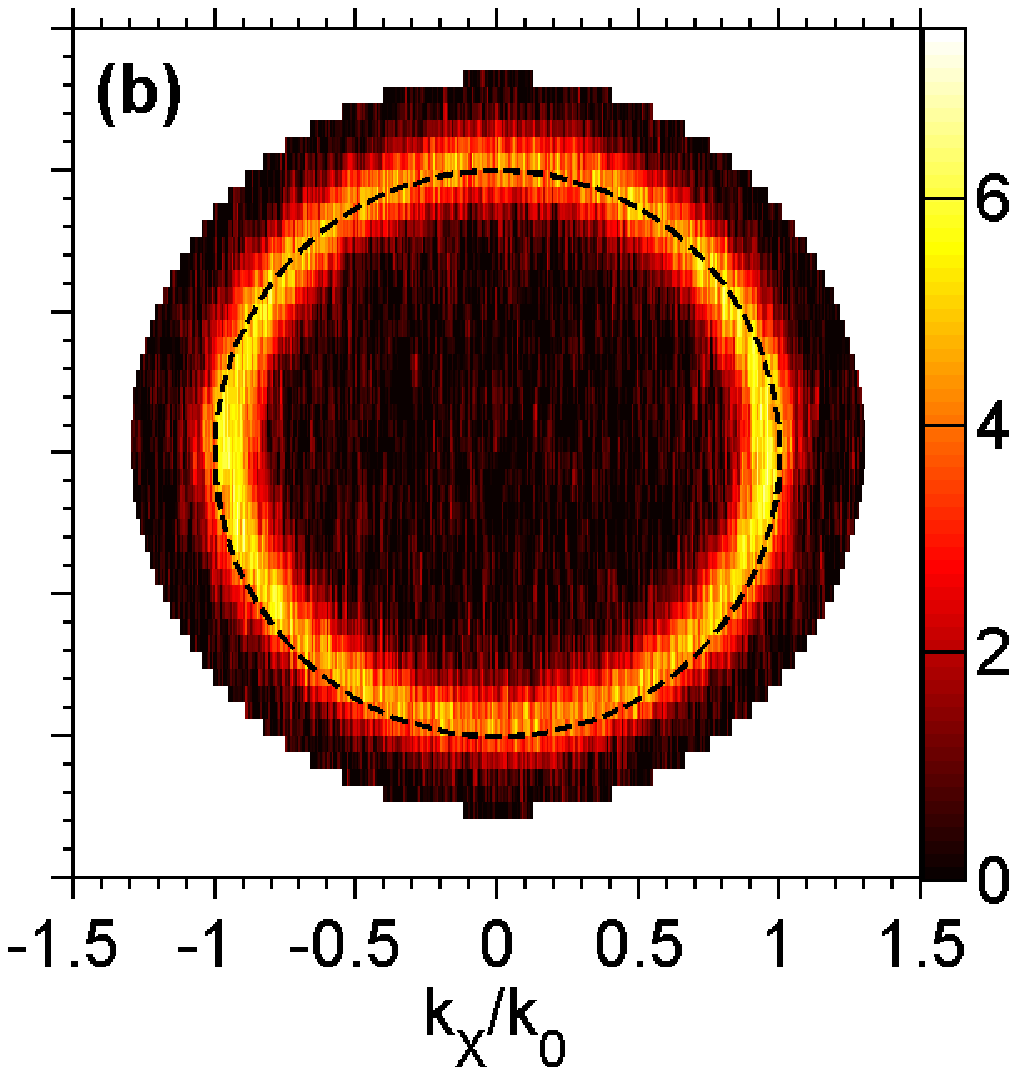} %
\includegraphics[width=8.2cm]{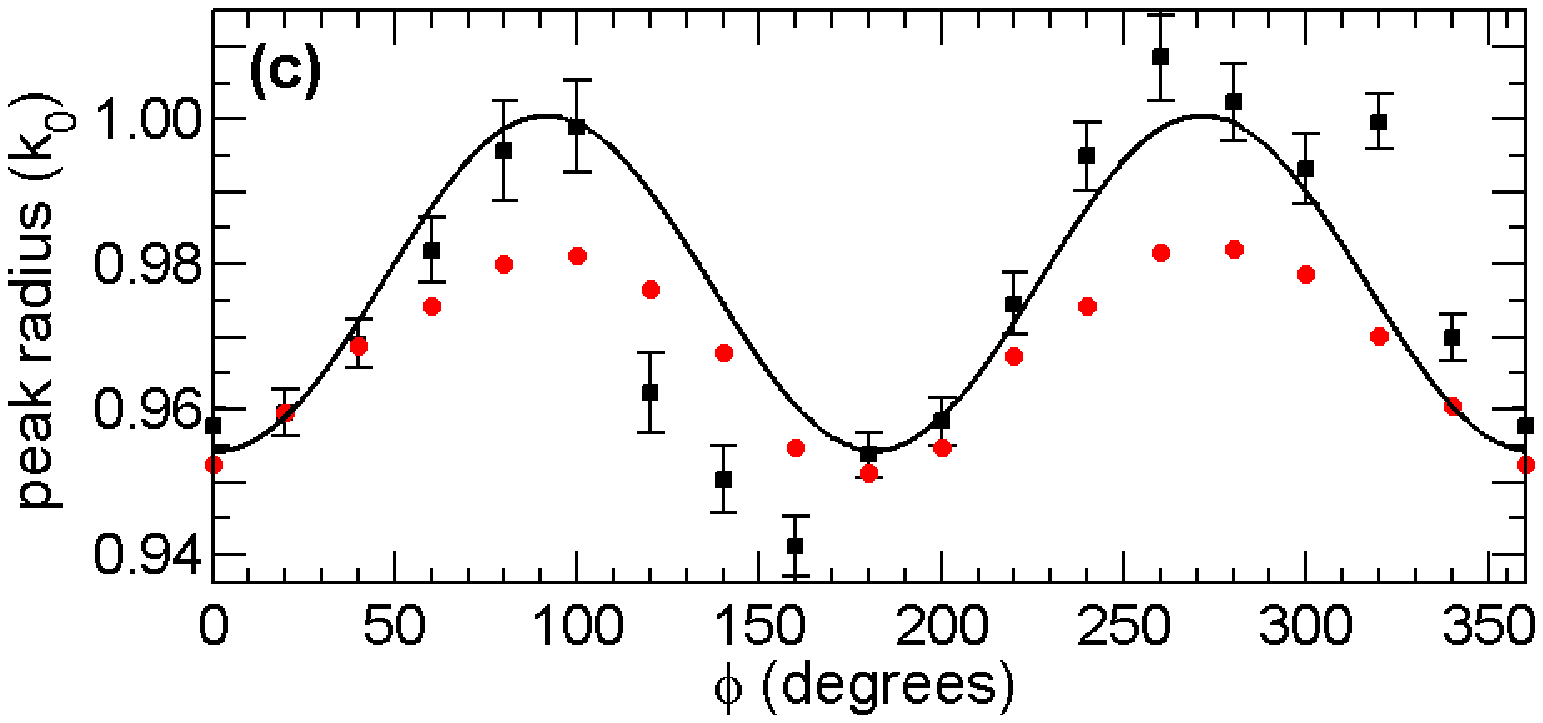}%
\end{center}
\caption{(Color online) (a) Average momentum space density $n(k_{X},k_{Y})$ (in arb. units,
from $\sim 1500$ experimental runs)
of the experimentally observed scattering halo on the equatorial
plane $(k_{X},k_{Y})$; the density is averaged over a disk of thickness
$[-0.1k_0,+0.1k_0]$ along $k_Z$. (b) Same as in (a) but from the
positive-$P$ simulation (see text) after $70$~$\mu$s
collision time, in units of $10^{-18}$ m$^3$. (c) Plot of the peak
radius of the scattering halo on the equatorial plane versus the
azimuthal angle $\protect\phi$.
Black squares are experimental data, while the red circles are
from the simulation.
The data is binned into 18 angular bins of $\Delta \protect\phi
=20^\circ$, and each data point for the peak radius is derived from
a Gaussian fit to the radial distribution
$n(k_R,\phi)\equiv n(k_X=k_R\cos \phi,k_Y=k_R \sin \phi)$
at the respective angle $\phi $
(the error bars show the statistical uncertainty in the fits; in
addition, there is a systematic uncertainty of $\pm1.5$\% in the
determination of the average radius of the sphere).
The smooth line is a sinusoidal fit to the experimental data. }
\label{fig:rawDensityData}
\end{figure}

In Fig.~\ref{fig:rawDensityData}~(a) we show a slice of the
scattering halo in the $XY$-plane that reveals its annular
structure. A dashed circle of radius 1, indicating the momentum
$\hbar k_0=mv_0$, is shown for comparison. We can see that the ring
corresponding to the mean momentum of scattered atoms does not lie
exactly on the dashed line, but rather slightly within it, and that
the deviation is anisotropic. The ring thickness and density
are also anisotropic, though in the present work we concentrate on
the behavior of the radius. 
To analyze the data more quantitatively,
we divide the ring into azimuthal sectors and fit a Gaussian peak plus a
linearly sloped background to extract a value for the halo radius as
a function of the angle $\phi$ ~\cite{perrin:08}. It is
clear from Fig.~\ref{fig:rawDensityData}~(c) that the radius of the
halo in momentum space varies approximately sinusoidally
by $\pm2\%$ and that it is almost always smaller than $k_0$.

%

To understand this result qualitatively, we first consider the
energy balance for pair production in a homogeneous BEC.
Removing an atom from the condensate liberates an energy
corresponding to the chemical potential, $g\rho$, where $g=4\pi
a\hbar ^{2}/m$, $a$ is the $s$-wave scattering length, and $\rho$
the density. Here, we have two counterpropagating condensates (each
having density $\rho/2$), which for simplicity we model as plane
waves. In the presence of the spatial modulation due to their interference,
the energy liberated by removing one atom changes to $3 g\rho/2$ \cite{simsarian:00}.
On the other hand, placing an atom in a scattering mode requires an energy
$2 g\rho$ since the scattered atom is distinguishable from those in
the condensate. Energy conservation, including the mean-field
contributions, gives
\begin{equation}\label{Ebalance}
\frac{\hbar ^{2}k_{0}^{2}}{2m}+\frac{3}{2} g\rho= \frac{\hbar
^{2}k_{s}^{2}}{2m}+2 g\rho,
\end{equation}
where we denote the absolute momentum of one scattered atom $\hbar
k_s$. Thus, the initial scattered momentum is \emph{smaller} than the
ingoing momentum, $k_{s}<k_{0}$. This effect was observed in a
numerical simulation in Ref.~\cite{bach:02}; a similar effect was
discussed in Ref.~\cite{vogels:02}. Using plane waves to model the
BECs is of course a crude approximation, but if we replace
$\rho$ by the central density of an inhomogeneous BEC, we
find $k_s=0.96\,k_0$ for the experimental parameters.

In addition to this \emph{initial} energy balance analysis, a second effect
must be taken into account. Once created, the scattered atoms
escape from the condensate region and gain energy from the
mean-field interaction potential. The effect is similar to that
reported in Ref.~\cite{simsarian:00}, an experiment which observed
the mutual repulsion of two BECs after Bragg diffraction.
If the source BEC were
stationary,
atoms would gain a kinetic energy $2g\rho$ as they
roll-off the mean-field potential. In our system however, the potential also
evolves in time and goes to zero in
the $XY$-plane on a timescale corresponding to the time for the two
condensates to separate ($\sim 70$ $\mu $s).
The rapid vanishing of the potential on the equatorial plane has a
very different effect on scattered atoms moving in the $X$ and $Y$
directions. Atoms moving along $Y$, the small dimension of the
trap, escape the condensate overlap region on a
timescale of $\sim 40$ $\mu $s, faster than the condensates can
separate. As a result, these atoms are accelerated by
a steep potential gradient and regain part of the energy $2g\rho $ (part -- because the potential itself is reduced during the separation).
On the other hand, atoms moving along
$X$, the long axis of the trap,
do not escape before the
condensates separate and thus experience much less acceleration.
Accordingly the observed momentum along the $X$ direction is smaller
than along $Y$, and much closer to the shifted value predicted by
Eq.~(\ref{Ebalance}).
%

To describe this experiment quantitatively we perform
first-principles positive-$P$ simulations similar to those in
Refs.~\cite{deuar:07,perrin:08}. Here, the
multimode dynamics of the atomic field operators $\hat{\Psi}(\mathbf{x},t)$ and $\hat{%
\Psi}^{\dagger }(\mathbf{x},t)$ for the $m_x=0$ state is fully modeled by
two independent complex $c$-fields, $\Psi (\mathbf{x},t)$ and $\tilde{\Psi}(%
\mathbf{x},t)$, satisfying the \^{I}to stochastic differential equations:%
\begin{eqnarray}
i\hbar \partial _{t}\Psi (\mathbf{x},t) &=&\mathcal{A}_{\mathrm{GP}}(\Psi ,%
\tilde{\Psi})\Psi +\sqrt{i\hbar g}\Psi\zeta _{1}(\mathbf{x},t),
\label{positive-P} \\
-i\hbar \partial _{t}\tilde{\Psi}(\mathbf{x},t) &=&\mathcal{A}_{\mathrm{GP}%
}(\Psi ,\tilde{\Psi})\tilde{\Psi}+\sqrt{-i\hbar g}\tilde{\Psi}\zeta _{2}(%
\mathbf{x},t).  \nonumber
\end{eqnarray}%
Here, $\mathcal{A}_{\mathrm{GP}}(\Psi ,\tilde{\Psi})=-\hbar
^{2}\nabla ^{2}/(2m)+g\tilde{\Psi}\Psi $ is a deterministic part
similar to the mean-field Gross-Pitaevskii (GP) equation, $\zeta
_{j}(\mathbf{x},t)$ ($j=1,2$) are
real independent noise sources with zero mean and correlations $%
\langle \zeta _{j}(\mathbf{x},t)\zeta _{k}(\mathbf{x}^{\prime
},t^{\prime })\rangle =\delta _{jk}\delta
^{(3)}(\mathbf{x}-\mathbf{x}^{\prime })\delta (t-t^{\prime })$,
while $g=4\pi \hbar ^{2}a/m$ uses $a=5.3$ nm \cite{perrin:07} for
the $m_x=0 $ atoms.

The initial condition for the outcoupled BEC in the $m_x=0$ state (assuming perfect outcoupling for simplicity) is a coherent state with the same density profile $\rho(%
\mathbf{x})$ as the trapped BEC in the $m_x=1$ state, with
$a=7.51$~nm \cite{moal:06}, $N_{0}=10^{5}$ atoms. Modulating this with
a standing wave imparts initial momenta $\pm k_{0}$ in the $Z$ direction,%
\begin{equation}
\Psi (\mathbf{x},0)= \langle \hat{\Psi}(\mathbf{x},0)\rangle =\sqrt{\rho(%
\mathbf{x})/2}\left( e^{ik_{0}\mathrm{Z}}+e^{-ik_{0}\mathrm{Z}}\right),
\label{initial-cond}
\end{equation}%
and models
the Bragg pulse that splits the BEC
into two equal halves described in the center-of-mass frame.
The initial density $\rho(\mathbf{x})$ is obtained as the ground state
solution to the
GP equation in the trap, and $\tilde{\Psi}(\mathbf{x},0) =\Psi(%
\mathbf{x},0)^*$.
The results of this simulation are shown in
Fig.~\ref{fig:rawDensityData}~(b) and (c) for $t=70$ $\mu$s at which
time the condensates have fully separated and the collision is over.
The result of the simulation is in reasonable agreement with the experiment.
The remaining discrepancy
could be because the experiment,
unlike the simulation, averages over a broad
distribution of initial atom numbers.
Since large condensates scatter more atoms, these events have more
statistical weight and bias the
data towards larger modulations.

\begin{figure}[tbp]
\includegraphics[width=8.2cm]{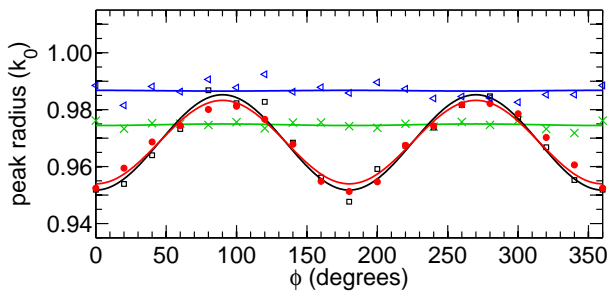}%
\caption{(Color online) Predictions for the peak radius of the scattering halo as in Fig.~\ref{fig:rawDensityData}~(c), after the end of the collision ($72$~$\mu$s),
with various controlled changes.
\emph{Red--$\bullet$}: full positive-P calculation, Eq.~(\ref{positive-P}) [same as in Fig.~\ref{fig:rawDensityData}~(c)];
\emph{Black--$\Box$}: anisotropic Bogoliubov calculation, Eq.~(\ref{STAB-equations});
\emph{Blue--$\lhd$}: anisotropic Bogoliubov, but with mean-field potentials $\propto g|\protect\psi_0|^2$ removed from Eq.~(\ref{STAB-equations})
and from the GP equation for $\psi_{0}(\mathbf{x},t)$;
\emph{Green--$\times$}: full Bogoliubov, but with \emph{spherical} BECs and unchanged peak density $\protect\rho(0)$
($200$~$\mu$s).
}
\label{fig:Bogoliubov}
\end{figure}

In order to confirm the qualitative mean-field mechanisms
described above,
we also perform an analysis of the collision dynamics using a
time-adaptive Bogoliubov approach \cite{STAB}, in which the atomic
field operator is split into the mean-field ($\psi_0$) and
fluctuating components, $\hat{\Psi}(\mathbf{x,}t)=\psi _{0}(%
\mathbf{x},t)+\hat{\delta}(\mathbf{x},t)$. The coherent BEC
wavefunction $\psi_0(\mathbf{x},t)$ evolves according to the
standard time-dependent GP
equation, with the initial condition given by Eq.~(\ref{initial-cond}).
The fluctuating component $\hat{\delta}(\mathbf{x},t)$ describes
incoherent scattered atoms, and is initially in the vacuum state.
In the
Bogoliubov approach, $\hat{\delta}$ evolves as
\begin{equation}\label{Bog-operator-eq}
i\hbar \partial _{t}\hat{\delta}(\mathbf{x},t)=%
\mathcal{H}_{0}(\mathbf{x},t)\hat{\delta}+\mathcal{G}(\mathbf{x},t)\hat{%
\delta}^{\dagger }.
\end{equation}
Here, $\mathcal{H}_{0}(\mathbf{x},t)=-\hbar ^{2}\nabla
^{2}/(2m)+2g|\psi _{0}(\mathbf{x},t)|^{2}$ contains the kinetic
energy and the mean-field potential energy $2g|\psi _{0}(\mathbf{x},t)|^{2}$
for scattered atoms.
The effective coupling $\mathcal{G}(\mathbf{x},t)=g\,\psi _{0}(\mathbf{x%
},t)^{2}$ causes spontaneous pair production of scattered atoms.
The dynamics of the field $\hat{\delta}$ is then formulated using
the positive-$P$ representation~\cite{STAB}, leading to the
(stochastic field) evolution equations
\begin{eqnarray}
i\hbar \partial_t\delta(\mathbf{x},t) &=& \mathcal{H}_0 \delta + \mathcal{G}%
\tilde{\delta} +\sqrt{i\mathcal{G}}\zeta _{1}(\mathbf{x},t),
\label{STAB-equations} \\
-i\hbar \partial_t\tilde{\delta}(\mathbf{x},t) &=& \mathcal{H}_0 \tilde{%
\delta} + \mathcal{G}^*\delta +\sqrt{-i\mathcal{G}^*}\zeta _{2}(\mathbf{x}%
,t),  \nonumber
\end{eqnarray}
which, unlike the full calculation~(\ref{positive-P}), are stable in
time because the noise is non-multiplicative. This method takes into
account the temporal evolution and spatial separation of the two
condensates; the stochastic formulation of the evolution of the
field $\hat{\delta}(\mathbf{x},t)$ makes explicit diagonalizations
on the (enormous) Hilbert space unnecessary.
As condensate depletion is $\sim 1.5\%$ here, the stochastic
Bogoliubov results are in excellent agreement with the positive-$P$
simulations, as seen in Fig.~\ref{fig:Bogoliubov}.

Figure~\ref{fig:Bogoliubov} also shows simulations performed
with controlled changes applied to the system.
The \emph{green} ($\times$) points use a spherical initial condensate
and show no anisotropy in the scattering sphere, unlike the \emph{black} ($\Box$) squares for the anisotropic case.
The \emph{blue} ($\lhd$) points have no mean-field potential, confirming that this
potential is essential for both the radius shift and the ellipticity.

%
%


The ability to detect three dimensional momentum vectors
of individual atoms allows the identification of small, previously unseen
anomalies in the scattering ``sphere'' resulting from a
simple collision between two condensates.
First-principles simulations
reproduce these small anomalies and help us to identify the important
physical processes.
An important application of pair production is the study and exploitation of quantum
correlations between the pairs, for example via Bell and
EPR type experiments \cite{RarityTapster,Kimble}. A matter-wave analogue of the optical EPR
experiment with parametric down-conversion \cite{Kimble} has been discussed
in Ref.~\cite{kheruntsyan:05} in the context of dissociation of a BEC of molecular dimers,
which produces atom-atom correlations similar to four-wave mixing.
In addition to the kinematic effects we report here, mean-field effects
will also affect the \emph{phases} of the associated two-particle wavefunctions.
Future work must carefully evaluate the effects of such (anisotropic and possibly fluctuating)
phase shifts on observables like the contrast of one- and
two-particle interference fringes.




\begin{acknowledgments}
We thank M. Gajda for valuable suggestions.
This work was supported by the French ANR, the IFRAF institute, and
the Euroquam Project CIGMA.
GP is supported by a European Union Marie Curie IIF Fellowship.
PD acknowledges the EU contract MEIF-CT-2006-041390.
PZ and MT are supported by Polish Government Research Grants.
KK acknowledges support by the Australian Research Council, and the hospitality of the Universit\'{e} Paris-Sud 11.
\end{acknowledgments}


\end{document}